\documentclass[a4paper,notitlepage,12pt]{article}
\usepackage{cite}

\newcommand\mks{\mbox{$\rm\mu s$}}%

\begin{document}

\title{Hypothesis of a daemon kernel of the Earth}

\author{E. M. Drobyshevski\\{\it Ioffe
Physical-Technical Institute, Russian Academy of Sciences,}\\{\it
194021 St.Petersburg, Russia. E-mail: emdrob@pop.ioffe.rssi.ru} }

\date{}

\maketitle

\begin{abstract}

The paper considers the fate of the electrically charged ($Z{\rm e} \approx  10{\rm e}$) Planckian elementary black holes -- daemons -- making up
the dark matter (DM) of the Galactic disk, which, as follows from our measurements, were trapped by the Earth during 4.5~Gyr in an amount of $\sim\!10^{24}$. Due to their huge mass ($\sim\!2 \times 10^{-8}$~kg), these
particles settle down to the Earth's center to form a kernel. Assuming the
excess flux of $10 \div 20$~TW over the heat flux level produced by known
sources, which is quoted by many researchers, to be due to the energy
liberated in the outer kernel layers in daemon-stimulated proton decay of
iron nuclei, we have come to the conclusion that the Earth's kernel is
presently measured in few fractions of meter in size. The observed mantle
flux of $\rm ^3He$ (and the limiting $\rm ^3He/\rm ^4He$ $\sim\! 10^{-4}$
ratio itself) can be provided if at least one $\rm ^3He$ (or $\rm ^3T$)
nucleus is emitted in a daemon-stimulated decay of $\sim\!10^2$--$10^3$~Fe nuclei.
This could remove actually the only objection to the hot origin of the Earth
and to its original melting. The high energy liberation at the center of the
Earth drives two-phase two-dimensional convection in its inner core (IC),
with rolls oriented along the rotation axis. This provides an explanation for
the numerous features in the IC structure revealed in the recent years
(anisotropy in the seismic wave propagation, the existence of small
irregularities, the strong damping of the P and S waves, ambiguities in the
measurements of the IC rotation rate etc.). The energy release in the kernel
grows continuously as the number of daemons in it increases. Therefore the
global tectonic activity, which had died out after the initial differentiation and cooling off of the Earth was reanimated $\sim\!2$~Gyr
ago by the rearrangement and enhancement of convection in the mantle as a
result of the increasing outward energy flow. It is pointed out that as the
kernel continues to grow, the tectonic activity will become intensified
rather than die out, as this was believed before.

{\ }

{\raggedright\it Keywords: dark matter; inner core; energy sources; convection;
helium; neon}

\end{abstract}

\nocite{*}

\section{Introduction. The problem of the energy source in the Earth and the
Dark Matter (DM) in the Solar system}

It appears obvious that global tectonics and generation of the Earth's
magnetic field are initiated by convective motions in its mantle and the
core. Recent estimates yield 44~TW for the global heat flux \cite{1,2,3}
. Radioactive decay of the lithophilic U, Th, and K, which are located
primarily in the crust, produces not more than 12--20 TW~\cite{1,2,3} .
 The nature of the other heat generators in the mantle and the core remains
unclear. The energy sources suggested besides the radioactive decay are tidal
friction, the continuing gravitational separation of material, the Earth's
cooling etc. However, they all are presently believed to be insufficient.
Lunar tidal friction produces now less than 1\% of the heat flux \cite{1}
. Gravitational energy of the separation could be substantial today too
\cite{1} , if one sticks to the classical "cold" (and slow) disk planetary
cosmogonies assuming very long accretion of the Earth ($\sim\!0.1$~Gyr and
even $\ge 1$~Gyr for the outer planets \cite{4} ), when its core melted out
of the cold material $\sim\! 2$~Gyr later, primarily due to radiogenic
heating of the interior. However, already the very first data on the Moon,
Mercury, Venus, and Mars obtained $in\  situ$ by space missions showed that all
these planets had been igneously differentiated possibly as early as in
the course of accretion, which assumes a hot protoplanetary nebula or a short
scale of their formation ($\sim\!10^4$--$10^5$~yr). It remains unclear how
could a high temperature ($\sim\!1000$--1500~K) in the disk, and the gaseous
disk itself (inevitably turbulized) persist for the required time of $\ge
0.1$~Gyr. The short
scale also can hardly be accounted for by nebular cosmogonies without
invoking additional hypotheses. It thus seems that a short and hot cosmogony
can be provided only by the hypothesis considering the Jupiter--Sun system as
the limiting case of a close binary star \cite{5,6}.

The acuteness of the problem of the inner Earth's heat is evident from the
recent review of Helfrich and Wood \cite{3} . They made in it, so to speak,
the last-ditch attempt at squeezing everything possible out of the
radioactive sources.

In formulating the problem, the authors point out that \cite{3}:

\noindent (1) The inventory of radioactive elements is capable at present of
producing 20~TW (i.e., about 45\% of the total heat flow);

\noindent (2) The content of $\rm ^{40}Ar$ in the atmosphere is only one half
that expected based on the assumed fraction of $\rm ^{40}K$ in the observed
heat balance;

\noindent (3) The $\rm ^4He$ flux from the oceans is only 5\% of the level
expected based on the oceanic heat flux.

Assuming that the outflow of heat resulting from the general cooling of the
Earth can reach 5.9--20~TW, Helfrich and Wood placed radioactive energy
release in (i) small ($<$4~km in size), seismically undetectable blobs of
subducted recycled parts of the oceanic crust distributed by overall
convection throughout the mantle, as well as in (ii) a near-core laterally
nonuniform $\rm D^{\prime \prime}$ layer (with an average thickness of $\sim\! 200$~km), where,
in their opinion, material of the oceanic crust could also build up. The
authors solve the problem of the deficiency of $\rm ^{40}Ar$ and $\rm ^4He$
evolving in the decay of K, U, and Th by invoking another hypothesis, namely,
they postulate that these gases are dissolved and accumulated in material
somewhere at the mantle base and, therefore, do not reach the Earth's
surface. No physicochemical arguments in favor of this assumption are
offered, so that nobody will be a penny the wiser! At the same time, it is
appropriate to note that the available isotope data argue (in accordance with
the opinion of Helfrich and Wood on the whole-mantle convection) for the
mantle being practically homogeneous chemically. Moreover, the $\rm ^3He / ^4He$ ratio
turns out to be maximum (up to $\sim\! 0.5 \times 10^{-4}$) in hot mantle
plumes (Hawaii, Iceland etc.) \cite {7,8}, which are believed (e.g. \cite{ 9,10}
 and refs. therein) to emanate exactly out of the lower mantle and maybe even
from the core \cite{11}.

Calderwood \cite{2} also carried out estimates of known sources of energy
liberation. These are radioactive heat production in the crust (9.13~TW),
lithosphere (0.38~TW), and in the depleted whole mantle (2.00~TW), as well as
the secular mantle cooling (10.0~TW). He comes to the conclusion that the
flux leaving the core reaches as high as $Q \ \sim\!21$~TW (!).

The above suggests that the missing source of energy and helium lies at a
depth $\ge 2700$~km, i.e., possibly, in the core, and reveals the complexity
of the problem facing the researcher and the need of looking for completely
new approaches to its solution \cite{11}.

It appears that the simplest and most natural mechanism of energy generation
in the interior of the Earth is suggested by our experiments \cite{12,13,14}
on the search for Dark Electric Matter Objects (daemons), i.e., elementary
Planckian black holes ($ r_{\rm g} = 3 \times 10^{-35}$~m, $m = 2 \times
10^{-8}$~kg), which make up the  DM of our Galaxy and of its disk. If they
carry a negative electric charge $Z\rm e=-10\rm e$, then their
accumulation in the Sun and catalysis of proton fusion there could account
not only for the deficiency of the $\rm ^8B$ neutrinos and the total absence
of $\rm ^7Be$ and $pep$ electron-capture neutrinos, but the solar energetics
itself \cite{15}.

Besides the possibility of catalysing light-nucleus fusion, the daemon, in
capturing a heavy nucleus and falling on the ground level located inside it,
brings about its overheating and evaporation of a part of the nucleons and of
their clusters out of the nucleus in $\sim\! 10^{-9}$~s, as well as gives
rise to a consecutive decay of its protons which it enters in the remnant of
the nucleus (each proton decay takes up $\sim\! 1$~$\mks $) \cite{14} (see
Sec.3).

A detector built with due account of the above properties of the daemon has
permitted us to record the flux of these particles, which cross the Earth's
surface, downward and upward, with an astronomic velocity of only $\sim\! 50$--5~$\rm km\,s^{-1}$, an unusually low figure for cosmic rays. The flux exhibits
strong seasonal variations and can be divided into several components,
namely, daemons moving with a velocity $V \sim\!$ 35--50~$\rm km\,s^{-1}$ make up the
population of the Galactic disk and/or of objects captured into strongly
elongated heliocentric orbits, objects with $V \approx 10(11.2)$ -- 15~$\rm km\,s^{-1}$
falling on the Earth from near-Earth, almost circular heliocentric orbits
(NEACHOs), and objects of the preceding group transferred into geocentric,
Earth-surface-crossing orbits (GESCOs), which constitute a population with a
velocity decreasing gradually from $\sim\! 10$ to $<$~5~$\rm km\,s^{-1}$ (by $\sim\!30$--
40\% per month) \cite{14}. Because of the daemons being slowed down by the
material of the Earth, these orbits contract gradually to finally become
confined within the Earth. Having a huge mass compared to the nuclei of
conventional elements (a difference of $\sim\!17$--19 orders of magnitude),
the daemons end up by reaching fairly rapidly a region near the Earth's
centre.

Markov \cite{16}, who started from the assumption that the daemons (which he
called "quantum maximons"), which build up inside the Earth and, in
transforming to conventional matter in interaction with one another, release
an energy $\sim\!m\,c^2$, estimated their flux through the Earth's surface as
$f_\oplus < 10^{-10}$~$\rm  m^{-2}\,s^{-1}$.

The purpose of this work is to consider, based on our measurements, some
implications of the accumulation of daemons inside the Earth, in particular,
the possibility of their accounting for the observed heat flux from the core.
This naturally suggests some conclusions concerning the properties of the
Earth's IC, which are found to fully agree with (and to be explained by) the
totality of the data available at present.

\section {The number of daemons inside the Earth}

By our measurements \cite{14}, the daemons fall on the Earth with a velocity
of $\sim\!$ 11.2--15~$\rm km\,s^{-1}$ apparently in the periods when it passes through
the "shadow"
or "antishadow" created by the Sun in the incoming flow of the Galactic disk
daemons. This occurs only during two months (approximately from
the end of February to the beginning of March, and, possibly, from the end of August to
 the beginning of September). This flux $f_\oplus \approx 20$~$\rm m^{-2}\,month^{-1}
 \approx 10^{-5}$~$\rm {m^{-2}\,s^{-1}}$. As already mentioned, part of this flux
is captured into GESCOs, with the velocity in these orbits decaying with time
as $\rm dln \it V /{\rm d}  t~=$~0.3--0.4~$\rm month^{-1}$ \cite{14}.

To estimate the slowing down of a daemon in one passage through the Earth,
assume that the GESCO period $P = 6000$~s. Then in each traversal of the
Earth, $2r_\oplus = 12750$~km, the daemon is decelerated by
$\Delta V = V_0P({\rm d}\ln V /{\rm d}t) = 11.2 \times 10^3 \:\times\: 6 \times
10^3\: \times\: 0.3\: /\: 2.5 \times 10^6 = 8$~m/s, which corresponds to a
decelerating force
$F \approx mV_0\Delta V/2r_\oplus  = 10^{-10}$~N.

If an object falling on the Earth with $V > V_0 = 11.2$ $\rm km\,s^{-1}$ crosses it with
$V \ge V_0$, it regains a NEACHO, i.e., it is not captured into a GESCO.
Thus, of the total number $\sim\! 3 \times 10^{16}$ daemons that fall on the
Earth's surface ($5.1 \times 10^{14}\  \rm m^2$) from
NEACHOs during one year (more specifically, during $\sim $ two months in a
year) only $3 \times 10^{16}\: \times\: 8\: /\: (15-11.2) \times 10^3 = 0.6 \times
10^{14}$ objects will be captured by the Earth into GESCOs (if they
annihilated, the heat flux from the Earth would exceed the observed figure by
three orders of magnitude). Assuming a constant flux, the Earth should have
acquired during its existence $3 \times 10^{23}$ negative daemons with a
total mass $\sim\! 10^{16}$~kg = 10~Tt, which, generally speaking, agrees
with an earlier estimate \cite{13} . Assuming for the sake of simplicity that
positive daemons also exist in equal numbers, and neglecting the differences
in the efficiency of their capture, the total mass will double to become
$M_{\rm k} \sim$ 20~Tt. Because our detector is to a certain extent
transparent to the daemons crossing it, we possibly slightly underestimate
this mass.

\section {Main conjectures on the state of daemons in the Earth and on their
interaction with matter}

In passing through the Earth's material, the GESCO daemons loose rapidly
their velocity, particularly if we recall that the force resisting the motion
of a charged particle increases generally with the decrease of its velocity
\cite{17}. Therefore the daemons end up near the Earth's center in a few
months. The loss in kinetic energy results in a release of energy, which is
less by $\sim\!$~six orders of magnitude than that from the core $Q$ (we
accept as a tentative estimate $Q = 10$~TW \cite{2,3}).

Our subsequent consideration will be based on the following simplifying
assumptions, some of which will be validated by the results obtained on their
basis:

\noindent(1) The daemons form at the Earth's center, inside the iron core
($\rho_0 = 1.3 \times 10^4$~${\rm kg/m^3}$, $p_0 = 360$~${\rm GPa}$
\cite{18}), a kernel of a practically collisionless daemon gas;

\noindent(2) As a consequence, the structure of the kernel can be
approximated by an isothermal ($T_{\rm d}~=~\rm const$), self-gravitating gas
sphere, so that the near-surface density $\rho_s$ of the daemon gas lies
within the limits $\rho_{\rm m} > \rho_{\rm s} > \rho_{\rm m}/3$ ($\rho_{\rm m}$
is the mean kernel density) \cite{19,20}. Because of the high pressure
generated  by the self-gravitation of the kernel at its center compared to
$p_0$, one may justifiably assume $\rho_{\rm s} = \rho_{\rm m}/3$;

\noindent(3) The pressure of the daemon gas at the sphere's boundary is
 balanced by that of the Earth's material, so that ($\rho_{\rm s}kT_{\rm d}/m
 \approx p_0 = 360$~GPa);

\noindent(4) Most of the energy 
is released in the successive disintegration of 
daemon-containing protons (938~MeV = $1.5 \times 10^{-10}$~J per 
event)
in
the Fe
nuclei captured by the daemon.
(In actual fact, this energy is liberated not in the nucleus itself
but in the surrounding material, wherefrom it is transported outward
by proton decay products, for instance, pions etc.).
The new nucleus is captured as soon
as the charge of the remnant of the preceding nucleus falls down to $Z - 1$,
making the effective daemon charge equal to $Z_{\rm eff}{\rm e} = -1 \rm e$.
In view of the fact that the binding energy of the nucleus to the daemon at
the ground level, when it resides inside a nucleus, is
$W~\approx~1.8ZZ_{\rm n}A^{-1/3}$~MeV, i.e., is measured in tens and hundreds
of MeV, one readily sees that such a repeated capture brings about a release
of such an energy that the remnant (say, oxygen, fluorine, or neon nucleus, depending on actual
value of $Z$) is ejected, and the anew captured, strongly excited Fe nucleus,
after
expending the excitation energy to evaporation, during
 $\tau_{\rm ev} \sim 10^{-9}$~s, of part of the nucleons ($\sim\!$ 8--10 in 
total)
 and/or of their
clusters
($\rm ^2D,\  ^3T,\  ^3He,\  ^4He,\  ^{10}Be$ etc; 
up to $\sim\!$ 5--6 $^4\rm He $ nuclei), retains $Z_n \le$ 14--22~protons
\cite{14}. From five to thirteen of them are subsequently disintegrated by the daemon
until the daemon/nuclear remnant system acquires a negative charge
 $Z_{\rm eff} = -1$. Therefore, on the average, the charge of the daemon
"poisoned" by the Fe nucleus is $Z_{\rm m} = $ 5--6. As follows from our
 measurements, the daemon-containing proton decays in $\Delta \tau \sim\! 1$~$\mks$,
 so that one cycle of recapture of a new iron nucleus lasts $\tau_{\rm ex}
\sim\!$ 10-11~$\mks \gg \tau_{\rm ev}$ \cite{14}. Therefore, each
negative daemon continuously "poisoned" by iron nuclei
({\it i}) creates, besides $\sim\! 10^6$
nucleons and their clusters, $\sim\!10^5$ nuclei of oxygen,
fluorine and neon per sec (in 4--10 sec, the $\rm\beta\mbox{-decay}$
 transforms $\rm ^{20-22}F$ into Ne),
and ({\it ii}) generates an energy
$q \approx 1.5 \times 10^{-4}$~J/s.

\section {Structure of the daemon kernel}

One can gain a certain idea of the possible parameters of the kernel, its
$\rho_{\rm m}$, $T_{\rm d}$, and radius $R_{\rm k}$, by assuming that the
energy flux leaving the IC is created mainly in the disintegration of protons in the
iron nuclei captured by daemons in the kernel.

It is unclear $a\  priori$ what part of the daemons takes part in this
process. We shall assume as a first approximation that these are objects
making up a surface layer of the kernel of thickness $l = (D\tau)^{1/2}$,
into which iron nuclei are capable of diffusing in a time
$\tau\  (\approx \ \tau_{\rm ex}\ \sim\!11$~$\mks)$. The diffusion coefficient of iron atoms (nuclei)
at temperature $T_{\rm Fe}$ will be defined as $D = \lambda V_{\rm T}/3$,
 $V_{\rm T} = (3kT_{\rm Fe} / m_{\rm Fe})^{1/2} \gg V_{\rm d}$, which is the
 daemon velocity in the kernel. The mean free path of iron nuclei
($Z_{\rm Fe} = 26$) in the near-surface plasma of "poisoned" daemons
($Z_{\rm m} = $ 5--6) can be written as \cite{17}.
$$\lambda = \frac{2\pi\varepsilon_0^2 m m_{\rm Fe}^2 V_{\rm
T}^4}{e^4Z_{\rm m}^2 Z_{\rm Fe}^2 \rho_{\rm s} \ln\Lambda}.$$
Here $\ln\Lambda \approx 10$ is the Coulomb logarithm, $\varepsilon_0 = 8.85
\times 10^{-12}$~F/m is the electrical constant.

Taking into account that the near-surface concentration $of\  
negative$ daemons is $\rho_{\rm s} / 2m$, we obtain
$$Q  = 4\pi R_{\rm k}^2 l q \rho_{\rm s} / 2m.$$
This equation has to be solved for $R_{\rm k}$ and $\rho_{\rm s}$ (for
$\rho_{\rm s} = \rho_{\rm m}/3$) together with
$$M_{\rm k}=4\pi{R_{\rm k}}^3\rho_{\rm s}.$$
Whence for $T_{\rm Fe} \approx 10^4$ K, $\ln \Lambda = 10$, and the other
above parameters
$$R_{\rm k}=0.048Q^2/M_{\rm k} \ \rm m;$$
$$\rho_{\rm s} = 700M_{\rm k}^4/Q^6 \ \rm Tt\,m^{-3},$$
where $Q$ is in TW, $M_{\rm k}$ is in Tt.

Then, at $Q = 10$~TW and $M_{\rm k} = 20$~Tt, $R_{\rm k} = 0.24$~m and
 $\rho_{\rm s} = 112$~$\rm Tt\,m^{-3}$.

One readily sees that the pressure $p_{\rm c}$ at the kernel's center
 generated by the self-gravitation of its material \cite{20}
$p_{\rm c} > \frac {3\pi}{8}\frac{GM_{\rm k}^2}{R_{\rm k}^4}$
 exceeds $p_0$ at its boundary by more than 12 orders of magnitude, so that
our approximation $\rho_{\rm s} = \rho_{\rm m}/3$ is fully justified. We may
 also recall that the density of atomic nuclear matter is $\approx 230$~$\rm
Tt\,m^{-3}$. Equating the pressures at the kernel's boundary $p = p_0 = 360$~${\rm GPa} = \rho_{\rm s}kT_{\rm d}/m$ yields the temperature of the
daemon
plasma $T_{\rm d}$ in the kernel $T_{\rm d} = 4.6 \times 10^9$~K, so that the kernel
daemons, because of their huge mass, move with a velocity of only
$V_{\rm d} \sim\!0.3$~cm/s. The high temperature of the kernel daemons 
(when $T_{\rm d} \gg T_{\rm Fe} $) is of
no surprise, because each proton disintegration releases 938~MeV 
while $m \gg m_{\rm Fe}$
so that an energy exchange between the components is strongly impeded.  

One should also note the following points:

\noindent(1) The thickness of the layer in which the energy is liberated is,
for our parameters, $l = 3 \times 10^{-8}$~m, which is an infinitesimal
fraction of the kernel dimensions while at the same time exceeding the mean
free path of the Fe nucleus here $\lambda = 1.4 \times 10^{-13}$~m. Thus, the
energy release is due to the activity of an extremely small ($ \sim\!2 \times
10^{-7}$) fraction of the accumulated daemons.

\noindent(2) The mean time between the Coulomb collisions of daemons with one
another for $\rho_{\rm m} = 3\rho_{\rm s}$ is estimated as \cite{17}.
$$\tau_{\rm dd} = \frac{2 \pi {\varepsilon_0}^2 m^3 V_{\rm d}^3}{0.714
\ln \Lambda \times 3 \rho_{\rm s} e^4 Z^4} \approx 10\ \rm s.$$

For a velocity $V_{\rm d} \sim\! 0.3$~cm/s, their mean free path will be
$\lambda_{\rm dd} = \tau_{\rm dd}V_{\rm d} \sim\! 3$~cm, which is comparable with the kernel dimensions. Therefore approximating the kernel by an isothermal
gaseous sphere is in this case justified (particularly if one takes
into account that the energy is liberated primarily near its surface). Whence
it follows, in particular, that the cross section of daemon Coulomb
interaction with one another is, on the average, only $\sigma_{\rm
dd}~=~(3\lambda_{\rm dd}n_{\rm s})^{-1}\rm \sim\!10^{-24}$~$\rm m^2$.

\noindent(3) Real physical collisions of daemons, when their centers approach
to a distance $ < 3 r_{\rm g}$, i.e., when one can, in principle, consider
their fusion and similar processes, are extremely rare. In our conditions
$\lambda_{\rm g} = (3n_{\rm s}9\pi{r_{\rm g}}^2)^{-1} \approx 10^{42}$~m, so
 that daemons collide with one another in the above manner once every
$\sim\!10^{45}$~s, i.e., once in every $\sim\!10^{21}$~s in the whole kernel.
In this sense, the daemon plasma is collisionless with a large margin.
Therefore there is hardly any sense in considering any noticeable energy
release in transformation of daemons in the Earth's kernel to some particles
or to conventional matter \cite{16}. It appears that this is valid for any
conceivable conditions. The only case in which physical collisions could
realize is possibly when the daemon ensemble escapes under the gravitational
radius (the case of quasars, active galaxy nuclei etc.(?)).

It appears more realistic to assume that the time required for daemons to
diffuse into the kernel is determined by the time $\tau=2\, \tau_{\rm ex}\,n_{\rm Fe}/n_{\rm s}$,
i.e., the time in
which each Fe nucleus of their total number $n_{\rm Fe} = \rho_{\rm 0}/m_{\rm Fe}$,
that had diffused from
the surface into the kernel, will be captured and processed by the
negative daemons present in the concentration $n_{\rm s}/2$. Then $R_{\rm k}=0.0051\, Q^{\rm 1/2}$~m,
and it will
be independent of $M_{\rm k}$ (so that the total energy release Q will be
proportional to the kernel surface area). For $Q = 10\ {\rm TW}$ and
$M_{\rm k} = 20\ {\rm Tt}$,
$R_{\rm k} = 1.6 \times 10^{-2}\ {\rm m}$, $T_{\rm d} = 1.4 \times 10^6\ {\rm K}$,
$\rho_{\rm s}=370 \times 10^3\ {\rm Tt\, m^{-3}}$,
$V_{\rm d} = 0.54 \times 10^{-4}\ {\rm m/s}$ and $l\approx 2 \times 10^{-9}\ {\rm m}$.

Despite the considerably smaller size of the kernel obtained under
these assumptions, the above arguments concerning its isothermality
and the collisionless conditions for the daemon population remain valid.

Further refinement of the kernel structure should be pursued,
in the first place, through a comprehensive analysis of the processes
involved in the energy and mass exchange between the kernel and
surrounding material of the Earth's inner core.

\section {Some inferences on the structure of the Earth's inner core}

The existence of a daemon kernel at the Earth's center and the associated
strong energy generation offer a possibility of understanding a number of the
following features of the Earth's IC revealed recently \cite{21}:

\noindent(1) It is usually believed that the IC is solid, first, because of a
density jump at its boundary (assuming the outer liquid core to have the same
composition), and, second, because the finite shear energy in its material
permits one to better reconcile the periods of normal oscillation modes of
its model with the observed periods \cite{22}. However, one cannot be
completely sure that the IC is fully solid \cite{23}. The calculations of Laio
$et\  al$. \cite{24} show that pure iron in ICB conditions is 6\% heavier than
the material of the outer core, which removes the first argument. The
observation of shear waves with the predicted velocity of 3.5--3.6~$\rm km\,s^{-1}$
\cite{25,26}  argues for the presence of solid material in the IC. However,
at frequencies $f > 0.1$~Hz they damp very strongly (with an amplitude
$\propto 10^{-13.6f}$ \cite{26,27}), which suggests an extremely nonuniform
structure of the IC and even the presence in it of a liquid phase
\cite{28,29}.

\noindent(2) There is an anisotropy in the propagation velocity of
longitudinal waves, with their velocity parallel to the Earth's axis of
rotation being 3-4\% larger than that parallel to the equator
\cite{30,31,32}.
Anisotropy can be simulated within an axisymmetric approach in the form of a
nonuniform IC having an inner central zone extended along the rotation axis,
which has a higher $P$-velocity \cite{27,33}. The most probable cause for the
anisotropy is believed to be preferential ($\sim\!1/3$) orientation of the
hexagonal close packed (hcp) crystals of the iron $\varepsilon$ phase
\cite{34} along the short basal-plane axes $a$ and $b$, the direction in
which the sound velocity exceeds by 12\% that along the long $c$ axis (for $T
= 5700$~K, the length ratio $a$:$b$:$c$ = 1:1:1.7) (see  \cite{35} and
refs. therein). Among causes for the orientation one put forward the magnetic
field-induced Maxwell stresses \cite{36,37} and structural changes in the
$\varepsilon$-Fe crystals caused by strain \cite{38} or their growth
(recrystallization) \cite{37,39} in the presence of solid state convection.

\noindent(3) Seismic data reveal also a substantial azimuthal asymmetry in
the IC structure. This refers both to the anisotropy, which on a large scale
is asymmetric relative to the rotation axis \cite{23} and seems to vary on a
time scale of tens of years \cite{21}, and to the presence of small-scale
($\sim\! 2$~km in the outer 300-km layer) irregularities \cite{28}. The
latter may originate from variations in the seismic anisotropy, material
composition, and even melt pods in the solid matrix. The irregularities do
not remain fixed with respect to observers on the Earth's surface, which gave
grounds to numerous reports of observation of differential IC rotation with a
rate of $\sim\! 1^{\circ}$ to $0.1^{\circ}$ per year, primarily (but not
always) in the easterly direction (see refs. in  \cite{21, 40}). These features
in the IC structure cause strong damping (or scattering) of the longitudinal
waves too.

It should be pointed out that although heat convection in the IC is believed
to be hardly likely (e.g. \cite{41}), nevertheless, all the above suggests
that some part (if only a small one) of the material is in the molten state
\cite{28} (everybody appears to believe that the temperature of the IC
material is close to the melting point $T_{\rm m}$).

The situation with interpretation of the above observations simplifies
considerably, if the IC contains a powerful source of energy $Q$, which
may drive a strong enough convection.

Convection sets in when the Rayleigh number
$$Ra = \frac{\alpha g Q R_{\rm ICB}^2}{4\pi\nu\chi^2\rho c_{\rm p}} \ge
Ra_{\rm c} \approx 3 \times 10^3.$$

In this case, if we take even the limiting values of the parameters quoted in
\cite{39} for a solid iron core of radius $R_{\rm ICB}$ at $T = 0.85 T_{\rm
m}$ [thermal expansion coefficient $\alpha \approx 3 \times 10^{-6}$~$\rm
K^{-1}$; kinematic viscosity $\nu = 10^{12}$~$\rm m^2\,s^{-1}$ (see also
 \cite{42}); thermal diffusivity $\chi = 2.5 \times 10^{-5}$~$\rm m^2\,s^{-
1}$
 (in our opinion, $\chi \sim\! 10^{-6}$~$\rm m^2\,s^{-1}$ would be more
reasonable); mass density $\rho = 13000$~$\rm kg\,m^{-3}$; specific heat
$c_{\rm p} = 500$~$\rm J\,kg^{-1}\,K^{-1}$; $g = 1$~$\rm m\,s^{-2}$], then
for $Q = 10$~TW we obtain $Ra = 0.8 \times 10^{11} \gg Ra_{\rm c}$, i.e.,
the system is unstable to convection with a huge margin.

We readily see that in these conditions the viscosity, whose magnitude is
very poorly known, exerts the largest effect on $Ra$.

For our values of $Q$, layers of a liquid phase are expected to form in the
solid convective phase (for instance, in zones of concentrated creep, where
(quasi)adiabatic shear belts are created), which, as we have seen, is in full
accord with observations of irregularities in the IC. Then the effective
viscosity of the system falls by many orders of magnitude. The Taylor number
$Ta = (2\Omega R_{\rm {ICB}}^2/\nu)^2$, which characterizes the ratio of the
Coriolis to viscous forces, i.e., the effect of rotation on convection, may
exceed its critical value $Ta_{\rm c} \approx$ 10--100. Convection becomes
two-dimensional and takes on the form of rolls parallel to the rotation axis
of the system \cite{43}. In our case, this happens when $\nu \le \nu_c
\approx 2\Omega{R_{\rm {ICB}}}^2/{Ta_{\rm c}}^{1/2} \cong 2.1 \times
10^7$~$\rm  m^2\,s^{-1} $. This value of $\nu_{\rm c}$ exceeds by far the viscosity
of molten iron ($\nu \approx 10^{-5}$~$ \rm m^2\,s^{-1}$) and even the lowest value $\nu \approx 10^6$~$ \rm m^2\,s^{-1}$
quoted in \cite{39} for solid iron at $T \approx 0.85 T_{\rm m}$. Therefore,
there are strong grounds to believe that rotation does indeed make convection
in the IC two-dimensional.

Obviously enough, the existence in the IC of two-phase, two-dimensional
convection with rolls oriented along the axis of rotation sheds light to a
certain extent on the nature of many of the above-mentioned observations.
Among them are, first, the presence of an elongated anisotropic zone inside
the IC, and, second, the nature of the anisotropy in the velocity of
longitudinal waves itself. Indeed, no assumptions on the strain or
recrystallization of the $\varepsilon$-Fe crystals are actually needed. The
fact (it can be checked by anybody in a tee cup) that elongated particles
acted upon by a viscous medium in shear flow are oriented along the flux
lines, which in the case of roll convection in a rotating system lie exactly
in planes normal to the rotation axis appears convincing enough.

The time-varying azimuthal irregularities (including small-scale ones) should
naturally appear in large-scale turbulent convection of a multi-phase medium,
with the solid-phase prevailing. The latter makes understandable the
generation and propagation in the IC of shear waves with a very strong
damping, which grows rapidly with a decrease in their length. If the
convection is of the roll character, probing the IC with seismic waves in
different directions and at different times may produce the impression of the
differential core rotation, with its measured velocity being different
depending on the actual conditions of observation.

\section {Conclusion}

The discovery of relic Planckian black holes with $m \sim\  2 \times 10^{-8}$~kg, which carry several electronic charges and make up the Dark Matter of the Galactic disk, leads to a number of far-reaching implications of a
fundamental nature. It appears that it offers us the long-awaited key, which
will provide a solution from the same standpoint for not only new but many
other, long-standing old problems, without invoking various $ad \  hoc$
 hypotheses. Among them is the explanation for the deficiency of the solar
neutrino flux and for the solar energetics itself, which has permitted us to
predict the decay of a daemon-containing proton and estimate its lifetime
\cite{44}. Our experiments on the search for daemons have confirmed this
estimate by yielding $\sim\! 1$~$\mks$ for the decay time \cite{14}.

Our experiments have revealed also a daemon population in NEACHOs and showed
that some of these objects are captured efficiently into geocentric,
Earth surface-crossing orbits. As a result, because of being slowed down by
the material of the Earth, they end up after a few months at its center and
build up there to have formed by this time a kernel $\sim\! 20$~Tt in mass
and $\le 1$~m in size. (These estimates can, of course, be revised somewhat
if the numbers of the positive and negative daemons in the Universe are
different or they have different efficiency of capture by the Earth.)

Despite its colossal density, the daemon matter of the kernel resides in the
state of a physically exotic, collisionless plasma. Indeed, because of the
infinitesimal cross section ($\sim\! 10^{-68}$~$\rm m^2$) and a low thermal
velocity ($V_{\rm d} \sim\  0.1$~m/s at a temperature of $\sim\! 5$~TK!),
direct collision of two daemons in the kernel occurs once in $\sim\! 1$~Tyr.
Actually, inside the Earth there is a laboratory for simulation of the
processes taking place in the Sun, as well as, on an immeasurably larger
scale, in quasars and active galactic nuclei! Nevertheless, it is clear that
pairwise interaction of daemons should not produce, as a rule, a noticeable
release of energy.

The existence of a daemon kernel removes a few well-known problems in the
physics of the Earth and, possibly, of other planets as well (e.g. excess heat
fluxes of the Jupiter, Saturn, and Neptune, generation of weak magnetic fields
of Mercury and Mars etc).

Primarily, this is the old problem of the energy balance of the Earth, which
consists in that known sources are capable of providing only $\sim\! 1/2$ of
the observed heat flux. It appears that the decay of daemon-containing iron
nuclei (more precisely, their protons) occurring in a thin outer layer of the
kernel provides a solution to
this problem. This also suggests reasons for convective motion in the outer
core, which is believed to be necessary for generation of the geomagnetic
field (the problem of the actual place of its generation needs now additional
studies).

The numerous problems revealed by the recent studies of the IC are likewise
solved. The presence of a powerful ($\sim\! 10$~TW) source of energy
inevitably gives rise to convection of the mixture of solid and liquid
material (iron). An analysis based on reasonable assumptions shows that the
rotation of the Earth organizes convective motions in the form of rolls
parallel to the rotation axis of the Earth. Orientation of elongated $\varepsilon$-Fe crystals by these motions accounts fully for the anisotropy
of seismic wave propagation in the IC. The existence of two-phase convection
explains the nature of small-scale irregularities in the IC, and its two-dimensional character can shed light on the reasons for the ambiguities in
the measurements of the differential IC rotation rate; maybe, it does not
rotate differentially at all, but the existence of convective rolls is
capable of simulating rotation, which is determined from the phase shift
between seismic waves passing through several rolls.
The existence in the Earth of a superdense kernel surrounded by a thick and
hot ( $T \geq 10^4 \, \rm K$ ) plasma envelope of an iron in the supercritical state may possibly
provide an explanation for the nature of some observations which
remain unclear (of the kind of the quasi-periodic movement of the Earth's
poles etc.), which apparently originate from the mobilities of the IC
\cite {40}, and now of the kernel, caused
by the tidal interactions in the Earth--Moon--Sun system.

The existence of a powerful unconventional nuclear source of heat under the
mantle clarifies
immediately the nature of the fixed hot spots and plumes on the Earth's
surface and provides also an explanation for the high $\rm ^3He/^4He$ ratio
found in them and the small cross section ($ < 100$~km) of the conjectured
ascending hot flows \cite {9}, which quite frequently escape detection by
seismic
methods. We may also recall the suggestion of the existence of superplumes
with thermal contrasts of up to $\sim\! 1000$~K, which was put forward more
than once \cite{9} but was never substantiated because of the absence of
known sources of energy under the mantle. The high $\rm ^3He/^4He$ ratio
itself (and even its absolute value of $\sim\! 10^{-4}$, but not larger) in
the plumes can be understood, if one assumes that the decay of
even one of
$\sim\!$ $10^2$--$10^3$ Fe nuclei excited in their capture by daemons is accompanied,
besides the emission of nucleons and their clusters of the type of $\rm ^4He$
and the like, by ejection of only one $\rm ^3He$ (or $\rm ^3T$)
nucleus. However, this is a rather low value, if we compare it with the reaction yield
$\rm ^3He(^3T)/^4He \approx 0.3$ for the case of Fe nuclei,
which become rapidly (and, possibly, even locally) overheated
by high-velocity particles (cosmic rays etc. \cite {45}).
The observed ratio $ \rm ^3He(^3T)/^4He \sim 10^{-4}$ can be accounted
for by the practically isothermal evaporation of nucleon clusters
from the Fe nucleus heated gradually to a nearly constant "boiling"
temperature by the daemon, which descends, step by step,
down to the ground level in the nucleus due to the stepwise
energy transfer to the most stable evaporating clusters of the type
of $\rm ^4He$.

There is also a certain possibility of the daemon capturing a $\rm ^4He$
nucleus, with its transformation after the proton decay to the radioactive
$\rm ^3T$ and, subsequently, to $\rm ^3He$. It is conceivable that
the daughter nuclei O, F, and Ne
likewise undergo similar transformations as well.

It appears that we have finally found the missing piece, that has been long
searched for by some researchers, "that nuclear reaction eluding the grasp
that is running (or was active in the past) in the interior of the Earth and
that produces a substantial amount of the $\rm ^3He$ daughter
isotope" \cite{46}.
Thus, we are witnessing erection of a real physical foundation for fairly
nonstandard conclusions considering the IC as a supplier of $\rm ^3He$,
which were drawn recently by Porcelly and Halliday \cite {11}
based on a comprehensive analysis of a fairly limited totality
of the presently available findings and observations.

Taking into account the role of the daemon kernel provides one more argument
for the fast (and hot) cosmogony of the Solar system, which considers it as
the limiting case of a close binary star \cite{5,6}, because it removes the
need of invoking the presently occurring (or recent), internal gravitational separation
of matter as the missing additional source of heat in the Earth. By the way,
production of $\rm ^3He$ (and, possibly, $\rm ^{2}D$, Ne, etc) at the
kernel/IC
interface removes apparently the only objection to the hypothesis of the
originally molten Earth, which consists in that the melt should have lost
rapidly the primary rare gases.

The significance of the presence of heat sources in the lower mantle and
under it for global tectonics is also obvious. It is usually assumed that the
Earth's heat decreases with time. And while in the Protoarchaean there
were many convective cells in the mantle, and the convective motions were
intense, which gave rise to formation of numerous small continents, the
gradual decrease in the heat flux brought about rearrangement and merging of
the cells and, accordingly, formation of large continents. This activity
decayed possibly to a minimum sometime $\sim\! 2$~Gyr ago \cite{47,48}, i.e.,
when the mass of the kernel in the Earth was one half its present value. One
might conjecture that tectonic activity has been growing thereafter
continuously (the formation and break up of the supercontinents Rodinia
($\sim\!$ 1000--750~Myr ago), Pannotia ($\sim\! 540$~Myr), Pangea
($\sim\!$200--150~Myr ago) \cite{48}) and will increase in intensity in the
future. Without questioning the "valve" role of individual continents in the
rearrangement of convective motions in the mantle in the present epoch \cite{10}, one can nevertheless predict that the steady growth of energy
liberation from the Earth's center will enhance convection, which will again
entail an increase in the number of cells and, as a consequence, in the
number of smaller floating continents.
The growth of liberated energy will be accompanied by the corresponding
flux increase of the simultaneously forming light isotopes
(including $ \rm ^3He$ and Ne).

\section*{Acknowledgements}

The author is greatly indebted to G.S.Anufriev and A.B.Mamyrin for fruitful
discussions on the $^3$He problem and related subjects. The main ideas of the
work were presented at the International Conference "AstroKazan--2001" (24--28~September, 2001, Kazan, Russia).

The work was partially supported by the Russian Foundation for Basic Research
(Grant 00-01-00482)

\end{document}